# ALL-IN meta-analysis for flexibility and validity in prospective and retrospective evidence synthesis

*invited contribution to [the special issue of the journal Cochrane Evidence Synthesis and Methods](), guest edited by Prof. Anna Lene Seidler and Dr. Peter Godolphin*
***Collaborative Evidence Synthesis: Individual Participant Data, Prospective Meta-analysis, and Other Approaches***

***Judith ter Schure,*** Corresponding author
Department of Epidemiology & Data Science, Amsterdam UMC, Amsterdam, The Netherlands.
Machine Learning Group, NWO-I institute CWI, Amsterdam, The Netherlands
https://orcid.org/0000-0002-2147-5510 j.a.terschure@amsterdamumc.nl

***Alain Amstutz***,
CLEAR Methods Center, Division of Clinical Epidemiology, Department Clinical Research, University Hospital Basel and University of Basel, Basel, Switzerland
Department of Research Support for Clinical Trials, Oslo University Hospital, Oslo, Norway.
Population Health Sciences, Bristol Medical School, University of Bristol, UK.
https://orcid.org/0000-0003-1716-993X alain.amstutz@unibas.ch

***Matthias Briel***,
CLEAR Methods Center, Division of Clinical Epidemiology, Department Clinical Research, University Hospital Basel and University of Basel, Basel, Switzerland.
https://orcid.org/0000-0002-2070-5230 matthias.briel@usb.ch

***Amir Aamodt Kazemi***,
Department of Research Support for Clinical Trials, Oslo University Hospital, Oslo, Norway.
Department of Rheumatology, Diakonhjemmet Hospital, Oslo, Norway.
Oslo Centre for Biostatistics and Epidemiology, Faculty of Medicine, University of Oslo, Oslo, Norway.
https://orcid.org/0000-0002-7778-6203 amirhosk@uio.no

***Inge Christoffer Olsen,***
Department of Research Support for Clinical Trials, Oslo University Hospital, Oslo, Norway.
https://orcid.org/0000-0001-6889-5873 i.c.olsen@medisin.uio.no

**Funding Statement**
This project was partially funded by the European Union's Horizon Europe research and innovation programme Grant Agreement No. 101156304 for PROACT EU RESPONSE.The funder had no role in the design, conduct, or publication of the content, research or evidence synthesis.

**CRediT authorship contribution statement**
*Judith ter Schure:* Conceptualization, Methodology, Data curation, Formal analysis, Visualization, Writing - original draft, Writing - review & editing
*Alain Amstutz:* Conceptualization, Writing - review & editing
*Matthias Briel:* Conceptualization, Funding acquisition, Writing - review & editing
*Amir Aamodt Kazemi:* Conceptualization, Writing - review & editing
*Inge Christoffer Olsen:* Conceptualization, Funding acquisition, Writing - review & editing

**Conflict of interest**
The authors declare no competing interests

**Ethics approval statement**
No ethics approval required.

**Patient consent statement**
No patient consent required.

**Permission to reproduce material from other sources**
Material reused from Replication Package ALL-IN-META-BCG-CORONA, available under CC-BY license: https://doi.org/10.53962/kyep-h9

Article type: commentary

**ABSTRACT (unstructured)**
ALL-IN meta-analysis was developed and first applied during the COVID-19 pandemic. While this setting inspired its name, ALL-IN can also benefit non-pandemic circumstances. Conventional meta-analysis loses its coverage when updated repeatedly over time and when the decisions to initiate new trials and synthesize them depend on the results within the meta-analysis (accumulation bias). ALL-IN meta-analysis is anytime-valid. In its simplest form, ALL-IN meta-analysis stays familiar to run and read based on forest plots with confidence intervals that are wider than standard ones. Within collaborative prospective meta-analysis, the payoff is flexibility and speed, with fast sharing of individual participant data or harmonized aggregate data. Outside of this setting, the payoff is validity, when the decisions that shape the evidence (when to stop trials and whether new ones start) are typically outside the control of the meta-analyst. ALL-IN meta-analysis becomes inefficient when there is a maximum sample size or a stopping rule that the meta-analyist can control. ALL-IN enables adaptations for any evidence synthesis to halfway become living, prospective or even real-time on interim trial results, without complicating the statistics.

## Introduction

ALL-IN meta-analysis is a statistical approach that is very promising for efficient, collaborative and adaptive evidence synthesis. In its simplest form it does one thing: it makes confidence intervals wider than in conventional meta-analysis to gain validity (conclusions drawn from accumulating evidence are more often right), and flexibility (those conclusions can be reached earlier, with a method that stays simple to run and read). ALL-IN stands for Anytime Live and Leading INterim meta-analysis (Ter Schure & Grünwald, 2025), and the name reflects its defining property: it is anytime-valid (Ramdas et al., 2023). Type-I error for tests and coverage for confidence intervals are guaranteed for unlimited updating, with no maximum sample size.

## Why conventional meta-analyses break under updating and ALL-IN does not

A conventional meta-analysis loses its coverage when updated repeatedly over time. The reason is the same multiplicity that inflates type-I error in clinical trials with repeated looks: Each update is another chance to be wrong, and across enough updates the true effect is eventually certain to fall outside a 95%-confidence interval. Pace and Salvan (2019) illustrated that a later confidence interval can completely contradict an earlier one when the two do not overlap – when this happens, at least one of them is wrong about the true effect. For standard 95%-confidence intervals used in meta-analysis, this happens more than 5% of the time, so the interval no longer delivers the confidence it claims.

ALL-IN meta-analysis is a sequential approach, like the alpha-spending methods in the Cochrane Handbook (Chapter 22.4; Higgins et al. 2024), but it is more flexible. Anytime-valid 95%-confidence intervals always have coverage and overlap of at least 95%, no matter how often they are updated. In the Handbook's terms, ALL-IN requires 'no pre-specified stopping rule'. How the meta-analysis is run and read can stay familiar. Figure 1 and 2 illustrate an ALL-IN meta-analysis (details below) in terms of the familiar forest plot (Figure 1) and an extension into a sequential forest plot (Figure 2), with wider, anytime-valid intervals.

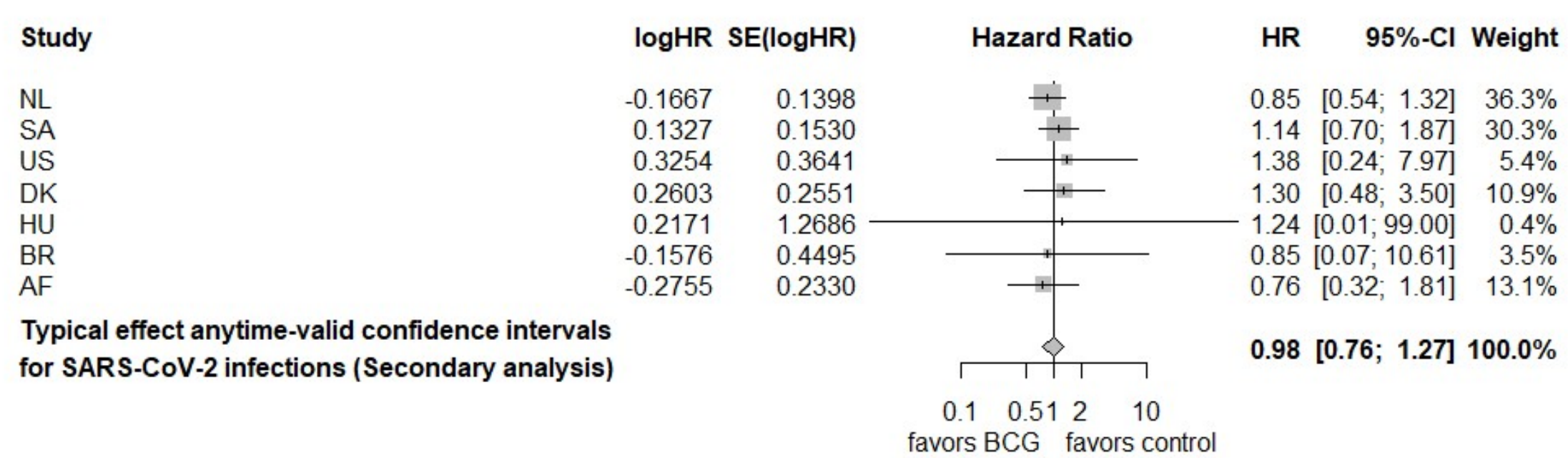

**Figure 1. ALL-IN meta-analysis forest plot of completed trials based on aggregate data.** Data from the secondary analysis of ALL-IN-META-BCG-CORONA with the seven trials from the 2022 synthesis (2026 update forthcoming with ten trials), see Ter Schure et al., 2022; specifically the Statistical Appendix to recalculate these. The anytime-valid confidence interval for the typical effect shown (fixed effects estimate) has a half-width of 3.037 times the SE(logHR) of 0.084: [0.76; 1.27] is $[\exp(\log(0.98) - 3.037*0.084); \exp(\log(0.98) + 3.037*0.084)]$, and 0.084 is $1/\sqrt{(1/0.1398^2 + 1/0.1530^2 + 1/0.3641^2 + 1/0.2551^2 + 1/1.2686^2 + 1/0.4495^2 + 1/0.2330^2)}$.
NL: the Netherlands (ClinicalTrials.gov: NCT04328441; Ten Doesschate et al., 2020, 2022; Claus et al., 2023; Van de Wijgert, 2024; Van Werkhoven et al., 2025), SA: South Africa (ClinicalTrials.gov: NCT04379336; Upton et al., 2022; van den Hoogen et al., 2026), US: United States (ClinicalTrials.gov: NCT04348370; Carrero Longlax et al., 2025; Mendez-Reyes et al., 2026), DK: Denmark (ClinicalTrials.gov: NCT04373291; Madsen et al. 2020, 2024; Nielsen et al., 2026), HU: Hungary (EU Clinical Trials Register: 2020-001783-28; Moldvay et al., 2026), BR: Brazil (ReBEC: RBR-4kjqtg; Junqueira-Kipnis et al., 2020; Borges Dos Anjos et al., 2022; Kipnis et al., 2026), AF: African trial in Guinea-Bissau and Mozambique (ClinicalTrials.gov: NCT04641858; Silva et al., 2025; Nielsen et al., 2026).
Plot based on summary statistics and R code available in replication package https://doi.org/10.53962/kyep-h9.

A standard 95%-confidence interval has a half-width of about two times the standard error (under normality assumptions: 1.96 times the standard error on both sides of the estimate). The anytime-valid intervals in Figure 1 and 2 have a half-width of about three times the standard error, at reasonable sample size. There is a large difference between the width at reasonable sample size and at small sample size. This is a feature, not a flaw: Like O'Brien-Fleming alpha spending and other procedures for repeated analysis, ALL-IN discourages strong conclusions in the early stages of data collection, yet it does so automatically through the width of the interval rather than through a separate rule.

## ALL-IN for pandemic and non-pandemic evidence synthesis

ALL-IN meta-analysis was developed and first applied during the COVID-19 pandemic (Ter Schure et al., 2022). While this setting inspired its name, ALL-IN meta-analysis can also benefit non-pandemic circumstances.

During a pandemic, the payoff is flexibility and speed. There is no maximum sample size, so the number of trials and participants can stay open-ended; interim analysis can be synthesized in real time, allowing highly effective treatments to be identified earlier without compromising validity. This mostly benefits collaborative prospective meta-analysis built on fast sharing of individual participant data or harmonized aggregate data.

Outside a pandemic, the payoff is validity. Here, the decisions that shape the evidence (when to stop trials and whether new ones start) are typically outside the control of the meta-analysist, both within trials – unplanned early stopping – and across them – accumulation bias (Ter Schure & Grünwald, 2019). We outline both these challenges in the next chapter and how ALL-IN remains valid regardless. As such, ALL-IN also applies to the standard, retrospective, non-collaborative, meta-analyses. We conclude with a discussion of limitations.

# 1. ALL-IN meta-analysis under pandemic circumstances

ALL-IN meta-analysis provides anytime-valid confidence intervals that can be interpreted at any time and updated indefinitely. This feature was crucial in the COVID-19 project where it was applied for the first time; the ALL-IN-META-BCG-CORONA collaboration (Ter Schure et al. 2022), see Figure 1 and 2. The common clinical question of more than ten very similar clinical trials was whether the BCG vaccine, originally developed to protect against tuberculosis, could also protect healthcare workers against SARS-CoV-2 infections. A prospective collaboration was established for dynamic and frequent meta-analyses. The idea was that by updating the ALL-IN meta-analysis often, a valid conclusion could be drawn early. An early conclusion of no clinically relevant benefit was available by the end of 2020. No trial had concluded by that time, but the ALL-IN meta-analysis could conclude based on two trials that at their interim stages captured evidence from an unforeseen wave of June-July winter infections in South-Africa, and an after-summer surge when restrictions were lifted in the Netherlands (SA and NL in Figure 2).

Confidence intervals for ALL-IN meta-analysis are wider, but because they allow for unplanned conclusions halfway trials and halfway evidence synthesis, the number of participants across all trials does not need to be much larger, on average. Here, ‘on average’ means that the conclusion can be reached much earlier, or much later, and we weigh all these scenarios together. This is a known result for other approaches to sequential analysis as well, like alpha spending (see a comparison in Figure 1 and 4 of Ter Schure et al., 2024).

In contrast to sequential analysis with alpha spending or a single meta-analysis, ALL-IN meta-analysis does not require a maximum sample size. This feature simplified the design of a collaboration like ALL-IN-META-BCG-CORONA. Defined in the number of infections, with the pandemic changing and the number of trials increasing, the accruing sample size was a truly open-ended process. Infections were driven by seasons and governmental pandemic restrictions, as is visible in the pattern (crosses for events of SARS-CoV-2 infections) at the bottom of Figure 2. Both recruitment of healthcare workers and infections were affected by surprising developments in the pandemic itself, such as the early rollout of COVID-19 specific vaccines of Pfizer and Moderna in some countries and almost total vaccine skepticism in others.

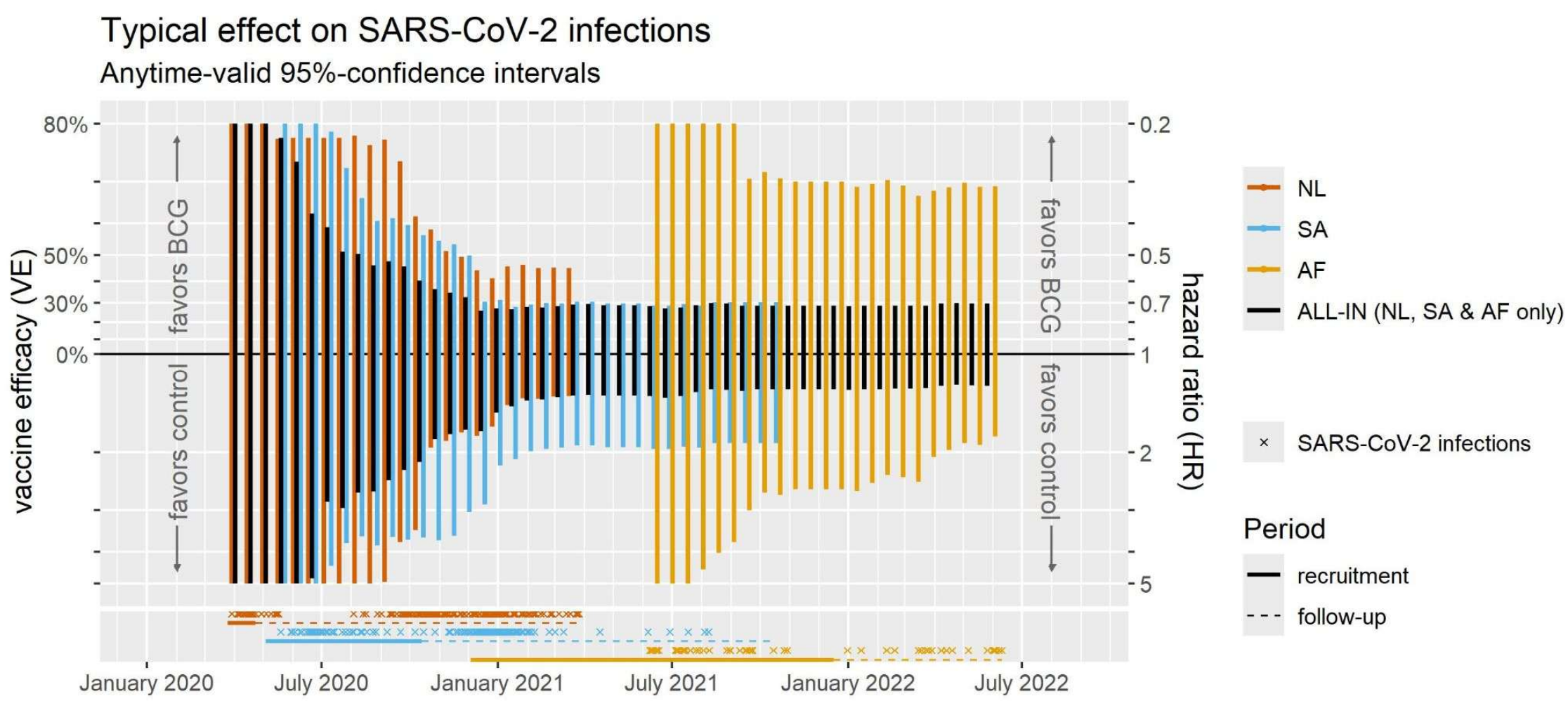

**Figure 2. ALL-IN meta-analysis sequential forest plot for real-time meta-analysis**. Data from the secondary analysis of ALL-IN-META-BCG-CORONA with only three trials for a less busy plot. These make up 80% of the total meta-analysis weight of the seven trials in the 2022 synthesis (2026 update forthcoming with ten trials), see Figure 1 and Ter Schure et al., 2022. Plot based on summary statistics and R code available in replication package https://doi.org/10.53962/kyep-h9.

# 2. ALL-IN meta-analysis outside of a pandemic

In single clinical trials, decisions to stop or continue at interim analyses are, or at least should be, described in the protocol. In the meta-analytic setting that control is usually impossible: the meta-analyst rarely governs when trials stop or whether new ones begin. ALL-IN delivers valid coverage even without it.

## 2.1 Stopping across trials: accumulation bias in any direction

Clinical trial science is cumulative: trial teams weigh existing results when deciding whether to initiate a new trial. This is good scientific conduct, yet it can bias a later meta-analysis, in either direction. The bias arises from decision-making over time and has been called accumulation bias (Ter Schure & Grünwald, 2019). Two examples show how.

### Non-conservative (upward) accumulation bias

The results of any clinical trial can be divided in four; a true positive or negative finding, or a false positive or negative finding. False negative trials could lead to decisions not to start a new trial because a conclusion is already reached and it would be considered a waste of resources (Chalmers and Glasziou, 2009) or even unethical to confirm harm on the primary outcome. Hence, the research question is stopped and the wrong conclusion is never rectified. On the other hand, a false positive finding could lead to interest in a field, and new trials could be initiated to further confirm the finding. Because the false positive trial then will be included in the following meta-analysis the point estimate will be biased against the null and lead to non-conservative (upward) accumulation bias. The meta-analysis only exists when more than one trial needs synthesizing which is conditional on the results.

### Conservative (downward) accumulation bias

In the case where two independent trials assess the same research question, but reach different conclusions, further trials are usually warranted to conclude. On the other hand, if the two reach the same conclusion, further trials may not be needed and a meta-analysis would also not add much that is not already known. In the first case the resulting meta-analysis will be biased towards the null and lead to a conservative bias because it by construct includes the false negative trial. Again, the meta-analysis only exists when more than two trials need synthesizing which is conditional on the results.

### Handling accumulation bias

These examples show that, unlike publication bias or p-hacking, accumulation bias arises exactly because science is working as it is supposed to. No data are hidden or distorted, and the best available evidence is used to decide what to run next (Lund et al., 2016). The bias is in which trials come to exist, and so in which meta-analysis we end up looking at. Only in some cases are the exact decision rules known and can the bias be quantified (see e.g. Pateras et al., 2012).

Accumulation bias arises from result-dependent stopping and continuation across trials: noisy results drive noisy decisions about whether the field accumulates one trial, two, or more. Anytime-valid intervals are guaranteed to cover the true effect under any such rule (known or unknown to the analyst), because their coverage holds at every update simultaneously, not only at one planned timing. The wider interval is precisely the price of this tolerance. Indeed, if we think of meta-analysis as a tool within cumulative science, any meta-analysis has the potential to be part of a sequential analysis.

## 2.2 Stopping within trials

Also within trials, decision-making takes place that affects the sample size. Sometimes based on planned and strictly enforced rules, but often not. Take a trial with a soft futility bound: the trial may stop on a disappointing interim (a true small effect plus unlucky noise) or continue when noise runs favourable, so stopped trials are biased downward (conservative) and the ones that continued are biased upward (benefit). For a specific trial that did not plan to enforce a strong rule, we can never know how large the bias is. The same applies to stopping for safety, or dropping the weaker arm in a multi-arm trial (Bauer et al., 2010).

Because ALL-IN is valid at every interim, it absorbs all of these scenarios. In our BCG vaccination example (Figure 1), five trials (US, DK, HU, BR, AF) stopped the recruitment of participants early. We cannot know for certain if the disappointing effect of the BCG vaccine influenced these decisions, but if it did, ALL-IN's validity holds.

# 3. Limitation of ALL-IN

ALL-IN's strength of adjusting its intervals to accommodate updates (both past and future), is also the source of its limitation. Securing valid 95% coverage under arbitrary updating costs interval width, and wider intervals mean less power to detect true effects (more type-II errors). In situations where there is nothing to update, and no uncontrolled decision-making to accommodate, that insurance goes unused and the extra width becomes an avoidable cost.

The planned prospective meta-analysis on treatment with tecovirimat in mild-to-moderate clade 2b mpox is such a case (Rojek et al, 2024). After mpox was declared a public health emergency of international concern in July 2022, five trials were launched under a common WHO CORE protocol (WHO, 2022), each independently of the other's result. With the outbreak resolved and two trials concluded (no clinically relevant effect), further trials on this treatment and population were unlikely. Here the set of trials is effectively closed and was never shaped by accumulating results. Conventional, unadjusted confidence intervals are valid, and ALL-IN would only make them needlessly wide.

# Discussion

ALL-IN meta-analysis delivers valid sequential inference (type-I error control and confidence interval coverage), much like the alpha-spending approaches discussed in the Cochrane Handbook Chapter 22.4 (Higgins et al., 2024). What sets it apart is that it needs no maximum sample size. A type-I error level (e.g. alpha = 5%, or 95%-confidence) is still fixed in advance, as well as a guess of the smallest effect size of interest tunes the sample size that is considered 'reasonable', but no 'stopping rule' is required. While excessive clinical trial research can be considered research waste if a research question is already answered or should not be prioritized (Chalmers & Glasziou, 2009), an ALL-IN meta-analysis does not meddle in these clinical considerations with statistical peculiarities. A forest plot or sequential forest plot can be judged on whether the synthesized interval shows relevant/irrelevant effects or is inconclusive and needs updating with more trials. No control over future or past trials is needed, and no redefinition of how the actual line of research came about in terms of an alpha-spending rule. Although not recommended in terms of research priority setting, new trials can be added indefinitely, and if we keep updating, we reach a conclusion – like the tests of power one that started the field of anytime-valid testing (Ramdas et al., 2023).

Bayesian methods are theoretically well-suited to both updating and accumulation bias (Cochrane handbook Higgins et al., 2024; Ter Schure & Grünwald, 2019; Chapter 5 Ter Schure, 2022), but the details are subtle. Convenient noninformative priors guarantee neither coverage nor 'Bayesian error control' (De Heide & Grünwald, 2021), and can also lose some of the Bayesian philosophy – as Stephen Senn (2011) captured in the article title *'You may believe you are a Bayesian but you are probably wrong'*. However, Bayesian meta-analysis as well as Bayesian adaptive trials have seen major developments, especially since the COVID-19 pandemic (Grant & Di Tanna, 2025; FDA draft guidance, 2026). Like ALL-IN, they share a conviction that accumulating knowledge demands more efficiency than fixed-sample methods allow. This philosophy is also central to collaborative evidence synthesis.

In conclusion, ALL-IN meta-analysis is a useful method for evidence synthesis wherever type-I error control and coverage matters and the analyses cannot be insulated from the dynamics of decision making within or across trials. The possibility of collaboration is always there, as ALL-IN allows adaptations for any evidence synthesis to halfway become living, prospective or even real-time on interim trial results, without complicating the statistics.

**Data availability**

All data from ALL-IN-META-BCG-CORONA presented here can be found in aggregate form by calendar date in a replication package available as a ResearchEquals collection: https://doi.org/10.53962/kyep-h9. R code to produce the forest plot and (simplified) sequential forest plot figures in this paper can be found there as well, https://doi.org/10.53962/pxav-kspy.

**Acknowledgements**

We acknowledge Declan Devane for nice discussions about this paper.

This project was partially funded by the European Union's Horizon Europe research and innovation programme Grant Agreement No. 101156304 for PROACT EU RESPONSE. The funder had no role in the design, conduct, or publication of the content, research or evidence synthesis.